\documentclass[prd,tightenlines,floatfix,showpacs,
nofootinbib,eqsecnum]{revtex4}
\usepackage[dvips,final]{graphicx}
\usepackage{epsfig}

\begin{document}

\date{\today}
\title{Impact of the higher twist effects on the $\gamma \gamma^{*} \to
\pi^{0}$ transition form factor }
\author{S. S. Agaev }
\email{agaev_shahin@yahoo.com}
\affiliation{Institut f\"ur Theoretische Physik, Universit\"at Regensburg,\\
D-93040 Regensburg, Germany\\
and\\
Institute for Physical Problems, Baku State University,\\
Z. Khalilov st.\ 23, Az-1148 Baku, Azerbaijan}

\begin{abstract}
We reanalize the $\gamma \gamma^{*} \to \pi^0$ transition form factor $%
F_{\pi \gamma}(Q^2)$ within the QCD light-cone sum rules method with the
twist-4 accuracy. In computations, the pion leading twist distribution
amplitude (DA) with two nonasymptotic terms and the renormalon
method-inspired twist-4 DAs are used. The latters allow us to estimate
impact of effects due to higher conformal spin components in the pion
twist-4 DAs on the form factor $F_{\pi \gamma}(Q^2)$. Obtained theoretical
predictions are employed to deduce constraints on the pion DAs from CELLO
and CLEO data.
\end{abstract}

\pacs{ 12.38.-t, 14.40.Aq, 13.40.Gp }
\maketitle

\section{Introduction}

The form factor (FF) $F_{\pi \gamma}(Q^2)$ of the electromagnetic transition 
$\gamma \gamma^{*} \to \pi^0$ is one of the simplest exclusive processes,
for investigation of which at large momentum transfers methods of the
perturbative QCD (PQCD) can be applied \cite{LB80,ER80,DM80}. For
computation of $F_{\pi \gamma}(Q^2)$ various theoretical methods and schemes
were proposed. They range from the PQCD calculations \cite{LB80,ABR83}, the
light-cone sum rules (LCSR) \cite{Kd99,SY00,BMS03}, to the running coupling
approach \cite{A01}, employed to estimate power-suppressed corrections to $%
F_{\pi \gamma}(Q^2)$. All of these methods are based on the PQCD
factorization theorems, in accordance of which the amplitude of an exclusive
process can be computed as the convolution integral of a hard-scattering
amplitude and the process-independent distribution amplitude (DA) of an
involved into the process hadron(s). The hard-scattering amplitude is
calculable within the QCD perturbation theory, whereas the hadron (in our
case, the pion) DAs are universal functions containing nonperturbative
information on hadronic binding effects, and cannot be obtained using tools
of PQCD. Hence, the hadron DAs emerge as one of the two building blocks in
studying numerous exclusive processes.

To calculate the $\gamma \gamma^{*} \to \pi^0$ transition FF in the
framework of the QCD LCSR method \cite{Kd99}, adopted also in this paper,
the knowledge of the pion different twist DAs $\varphi_{\pi}^{(t)}(u,
\mu_{F}^2)$ is required. As we have just mentioned, they cannot be found by
means of PQCD. Their factorization scale $\mu_{F}^2$ dependence is governed
by PQCD, but an input information at the starting point of evolution, i.e.,
the dependence of the DAs on the variable $u$ (the longitudinal momentum
fraction carried by the quark in the pion) at the normalization point $%
\mu_0^2$, has to be extracted from experimental data or derived via
nonperturbative methods, for example, the QCD sum rules \cite{CZ84},
instanton-based models \cite{Dor}, or lattice simulations \cite{Lattice}.
Nevertheless, there exists the regular theoretical approach for treatment of
the hadron DAs. It suggests the parametrization of the hadron DAs in terms
of a partial wave expansion in conformal spin, and rely on the conformal
symmetry of the QCD Lagrangian \cite{BKM03}. It is important that any
parametrization of DA based on a truncated conformal expansion is consistent
with the QCD equations of motion \cite{BF90} and is preserved by the QCD
evolution to the leading logarithmic accuracy \cite{ER80,LB80}. Therefore,
the conformal expansion provides a practical framework for modeling of the
hadron DAs \cite{BF90,BB98} and is widely used for investigation of numerous
exclusive processes in QCD.

Because of the increasing number of parameters at higher conformal spins and
practical difficulties in phenomenological applications, one has to restrict
one's self by taking into account only the first few terms in the conformal
expansion of DAs. As a result, the contributions of higher conformal spins
to DAs in the existing calculations are neglected. At the same time, the
suppression of higher spin contributions and the convergence of conformal
expansion at present experimentally accessible energy regimes is not obvious
and requires further studies.

The renormalon model proposed in Refs.\ \cite{An00,BGG04} pursues to test
precisely this issue; that is, to set a plausible upper bound for the
possible contributions of higher conformal spins that so far escaped
attention. The renormalon approach employs the assumption that the infrared
renormalon ambiguities in the leading twist coefficient functions should
cancel the ultraviolet renormalon ambiguities in the matrix elements of
twist-4 operators in a relevant operator product expansion. Such
cancellation was proved by explicit calculations in the case of the simple
exclusive amplitude involving pseudoscalar and vector mesons \cite{BGG04}.
The idea of the renormalon model for the meson twist-4 DAs is to define them
by taking the functional form of the corresponding ultraviolet renormalon
ambiguities and replacing the overall normalization constant by a suitable
nonperturbative parameter. It turns out that this is enough to obtain the
set of two- and three-particle twist-4 DAs of the pion and and $\rho$-meson
in terms of the corresponding leading twist (twist-2) DAs. It is remarkable
that the set of twist-4 DAs, apart from the parameters in the leading twist
DA, depend only on one new parameter. The latter can be related to the
matrix element of some local operator and estimated using the QCD sum rules.

A generic feature of the renormalon model is that it predicts higher twist
distributions that are larger at the end points compared to the lowest
conformal spin (i.e., the asymptotic) distributions, and are expected to
modify a behavior of higher twist contributions in exclusive reactions. In
fact, in our previous work \cite{A02} we employed the renormalon-inspired
twist-4 DAs for computation of the pion electromagnetic FF $F_{\pi}(Q^2)$ in
the context of the LCSR method, and found that the new DAs enhance the
twist-4 component of FF starting from $Q^2 \geq 5.5\ \mathrm{{GeV}^2}$, and
shift it towards larger values of $Q^2$. Such modification affects the data
fitting procedure and extraction of the parameters $b_2(\mu_0^2),\
b_4(\mu_0^2)$ in the pion leading twist DA, because in the renormalon
approach the twist-4 contribution to $F_{\pi}(Q^2)$ depends on the same
parameters, as the twist-2 one, and is not a "frozen" background like in the
standard analyses \cite{BH94}.

In the present work we reanalyze the $\gamma \gamma^{*} \to \pi^0$
transition FF within the QCD LCSR method applying the renormalon-inspired
model for the twist-4 DAs. We compare our predictions with the CELLO \cite%
{CELLO} and CLEO \cite{CLEO} data on this process and deduce constraints on
the parameters $b_2(\mu_0^2),\ b_4(\mu_0^2)$ at the normalization scale $%
\mu_0^2=1\ \mathrm{{GeV}^2}$.

This work is organized as follows: In Sec.\ II we define the two-- and
three-particle twist-4 DAs of the pion relevant to our present consideration
and introduce their models in the renormalon approach. In Sec.\ III general
expressions for the FF $F_{\pi \gamma }(Q^{2})$ in the QCD LCSR method, as
well as our results for the twist-4 contribution, are presented. In Sec.\ IV
we compare our predictions with the CELLO and CLEO data on the $\gamma
\gamma^{*} \to \pi^0$ transition and obtain constraints on the parameters $%
b_2(\mu_0^2),\ b_4(\mu_0^2)$. Section\ V contains our conclusions. Some
important but cumbersome expressions are collected in the Appendix.

\section{The renormalon model for the pion DAs}

In general, a pion is characterized by distributions of different partonic
contents and twists. Its leading twist DA corresponds to a partonic
configuration of the pion with a minimal number (quark-antiquark) of
constituents. But the light-cone expansion of the relevant matrix element
gives rise to two--particle higher twist DAs as well. The parton
configurations with a nonminimal number of constituents (for example,
quark-antiquark-gluon) are another source of the pion higher twist DAs. We
concentrate here only on DAs that will be used later in our calculations.

The light-cone two-particle DAs of the pion are defined through the
light-cone expansion of the matrix element, 
\[
\left\langle \pi^0(p) \left| \overline{u}(x_{2})\gamma _{\nu }\gamma _{5}%
\left[ x_{2},x_{1}\right] u(x_{1})\right| 0\right\rangle = 
\]
\[
=-i\frac{f_{\pi }p_{\nu }}{\sqrt{2}}\int_{0}^{1}due^{-iupx_{1}-i\overline{u}%
px_{2}}\left[ \varphi^{(2)}(u,\,\mu _{F}^{2})+\Delta ^{2}\varphi
_{1}^{(4)}(u,\,\,\mu _{F}^{2})+O(\Delta ^{4})\right] 
\]
\begin{equation}
-i\frac{f_{\pi }}{\sqrt{2}}\left( \Delta _{\nu }(p\Delta )-p_{\nu }\Delta
^{2}\right) \int_{0}^{1}due^{-iupx_{1}-i\overline{u}px_{2}}\left[ \varphi
_{2}^{(4)}(u,\,\,\mu _{F}^{2})+O(\Delta ^{4})\right] .  \label{eq:2.1}
\end{equation}%
Here $\varphi ^{(2)}(u,\,\mu _{F}^{2})\equiv \varphi _{\pi }(u,\,\mu
_{F}^{2})$ is the leading twist DA of the pion, and $\varphi
_{1}^{(4)}(u,\,\,\mu _{F}^{2}),\,\,\varphi _{2}^{(4)}(u,\,\,\mu _{F}^{2})$
are its two-particle twist-4 DAs. We use the notation $\left[ x_{2},x_{1}%
\right] $ for the Wilson line connecting the points $x_{1}$ and $x_{2}$: 
\begin{equation}
\left[ x_{2},x_{1}\right] =P\exp \left[ -ig\int_{0}^{1}dt\Delta _{\mu
}A^{\mu }(x_{2}+t\Delta )\right] .  \label{eq:2.2}
\end{equation}
In Eqs.\ (\ref{eq:2.1}) and (\ref{eq:2.2}) $\Delta =x_{1}-x_{2}$ and$\,\, 
\overline{u}=1-u$.

The three-particle twist-4 DAs involving an extra gluon field can be
introduced in the form \cite{BF90} 
\[
\left\langle \pi^0(p)\left| \overline{u}(-z)\left[ -z,vz\right] \gamma _{\nu
}\gamma _{5}gG_{\mu \rho }(vz)\left[ vz,z\right] u(z)\right| 0 \right\rangle 
\]
\[
=\frac{f_{\pi }}{\sqrt{2}}\int D\alpha _{i}e^{-ipz(\alpha _{1}-\alpha
_{2}+\alpha _{3}v)}\left\{ \frac{p_{\nu }}{pz}(p_{\mu }z_{\rho }-p_{\rho
}z_{\mu })\Phi _{\parallel }(\alpha _{1},\alpha _{2},\alpha _{3})\right. 
\]
\begin{equation}
\left. +\left[ p_{\rho }\left( g_{\mu \nu }-\frac{z_{\mu }p_{\nu }}{pz}
\right) -p_{\mu }\left( g_{\rho \nu }-\frac{z_{\rho }p_{\nu }}{pz}\right) %
\right] \Phi _{\perp }(\alpha _{1},\alpha _{2},\alpha _{3})\right\} ,
\label{eq:2.3}
\end{equation}
where the longitudinal momentum fraction of the gluon is $\alpha _{3}$ and
the integration measure is defined as 
\begin{equation}
\int D\alpha _{i}=\int_{0}^{1}d\alpha _{1}d\alpha _{2}d\alpha _{3}\delta
(1-\alpha _{1}-\alpha _{2}-\alpha _{3}).  \label{eq:2.4}
\end{equation}
The other pair of DAs is obtainable from Eq.\ (\ref{eq:2.3}) after the
replacement $\gamma _{5}G_{\mu \rho }\to i\widetilde{G}^{\mu \rho }= \frac{i%
}{2}\epsilon ^{\mu \rho \alpha \beta }G_{\alpha \beta },$ 
\[
\left\langle \pi^0(p)\left| \overline{u}(-z)\left[ -z,vz\right] \gamma _{\nu
}ig \widetilde{G}_{\mu \rho }(vz)\left[ vz,z\right] u(z)\right| 0
\right\rangle 
\]%
\[
=\frac{f_{\pi }}{\sqrt{2}}\int D\alpha _{i}e^{-ipz(\alpha _{1}-\alpha
_{2}+\alpha _{3}v)}\left\{ \frac{p_{\nu }}{pz}(p_{\mu }z_{\rho }-p_{\rho
}z_{\mu })\Psi _{\parallel }(\alpha _{1},\alpha _{2},\alpha _{3})\right. 
\]%
\begin{equation}
\left. +\left[ p_{\rho }\left( g_{\mu \nu }-\frac{z_{\mu }p_{\nu }}{pz}
\right) -p_{\mu }\left( g_{\rho \nu }-\frac{z_{\rho }p_{\nu }}{pz}\right) %
\right] \Psi _{\perp }(\alpha _{1},\alpha _{2},\alpha _{3})\right\}.
\label{eq:2.5}
\end{equation}

There exist one more three-particle twist-4 DA $\Xi _{\pi }(\alpha _{i})$ %
\cite{BGG04}, as well as, four quark twist-4 distributions, which we do not
consider in this paper.

The pion two-- and three-particle twist-4 DAs are not independent functions,
because the QCD equations of motion connect them with each other. From the
analysis based on exact operator identities \cite{BF90,Br89}, it follows
that 
\[
\varphi _{2}^{(4)}(u)=\int_{0}^{u}dv\int_{0}^{v}d\alpha
_{1}\int_{0}^{1-v}d\alpha _{2}\frac{1}{\alpha _{3}}\left[ 2\Phi _{\perp
}-\Phi _{\parallel }\right] (\alpha _{1},\alpha _{2},\alpha _{3}), 
\]%
\begin{equation}
\varphi _{1}^{(4)}(u)+\varphi _{2}^{(4)}(u)=\frac{1}{2}\int_{0}^{u}d\alpha
_{1}\int_{0}^{1-u}d\alpha _{2}\frac{\overline{u}\alpha _{1}-u\alpha _{2}}{%
\alpha _{3}^{2}}\left[ 2\Phi _{\perp }-\Phi _{\parallel }\right] (\alpha
_{1},\alpha _{2},\alpha _{3}),  \label{eq:2.6}
\end{equation}%
where $\alpha _{3}=1-\alpha _{1}-\alpha _{2}$.

The renormalon method provides the new, additional relations between
different twist DAs of the pion. In order to explain principle points of the
renormalon approach and derive relations between the pion twist two and four
DAs in Ref.\ \cite{BGG04}, the authors considered the gauge-invariant
time-ordered product of two quark currents, 
\[
\left\langle 0\left| T\{ \overline{d}(x_{2})\gamma _{\nu }\gamma _{5}\left[
x_{2},x_{1}\right] u(x_{1})\}\right| \pi ^{+}(p)\right\rangle, 
\]
at small light-cone separations and expressed the matrix element in terms of
two Lorentz-invariant amplitudes $G_i(u, \Delta^2),\ i=1,2$. They applied
the operator product expansion to the amplitudes $G_i(u, \Delta^2)$,
computed the infrared renormalon ambiguities of the twist-2 coefficient
functions and ultraviolet renormalon ambiguities arising from higher twist
operators, and proved that these ambiguities cancel exactly in OPE,
rendering the structure functions $G_i(u, \Delta^2)$ unambiguous to the
twist-4 accuracy. In the renormalon model, one defines the pion twist-4 DAs
by keeping the functional form of the corresponding ultraviolet renormalon
ambiguities and replacing the overall normalization constant $c\Lambda^2$ by
the nonperturbative parameter $\delta^2/6$. We refer the readers to Ref.\ %
\cite{BGG04} for detailed analysis and calculations, and write down only
final results: 
\[
\Phi _{\perp }(\alpha _{1},\alpha _{2},\alpha _{3})=\frac{\delta ^{2}}{6}%
\left[ \frac{\varphi _{\pi }(\alpha _{1})}{1-\alpha _{1}}-\frac{\varphi
_{\pi }(\alpha _{2})}{1-\alpha _{2}}\right], 
\]
\[
\Phi _{\parallel }(\alpha _{1},\alpha _{2},\alpha _{3})=\frac{\delta ^{2}}{3}
\left[ \frac{\alpha _{2}\varphi _{\pi }(\alpha _{1})}{(1-\alpha _{1})^{2}}- 
\frac{\alpha _{1}\varphi _{\pi }(\alpha _{2})}{(1-\alpha _{2})^{2}}\right], 
\]
\[
\Psi _{\perp }(\alpha _{1},\alpha _{2},\alpha _{3})=\frac{\delta ^{2}}{6}%
\left[ \frac{\varphi _{\pi }(\alpha _{1})}{1-\alpha _{1}}+\frac{\varphi
_{\pi }(\alpha _{2})}{1-\alpha _{2}}\right], 
\]
\begin{equation}
\Psi _{\parallel }(\alpha _{1},\alpha _{2},\alpha _{3})=-\frac{\delta ^{2}}{%
3 }\left[ \frac{\alpha _{2}\varphi _{\pi }(\alpha _{1})}{(1-\alpha _{1})^{2}}%
+ \frac{\alpha _{1}\varphi _{\pi }(\alpha _{2})}{(1-\alpha _{2})^{2}}\right].
\label{eq:2.7}
\end{equation}
In the left-hand sides of Eq.\ (\ref{eq:2.7}) the substitution $%
\alpha_3=1-\alpha_1-\alpha_2$ is implied.

Having substituted Eq.\ (\ref{eq:2.7}) into Eq.\ (\ref{eq:2.6}) and computed
the relevant integrals, one can obtain the pion two-particle twist-4 DAs $%
\varphi_1^{(4)}(u,\mu_F^2)$ and $\varphi_2^{(4)}(u,\mu_F^2)$ in terms of the
leading twist DA. Stated differently, the twist-4 DAs (\ref{eq:2.1}), (\ref%
{eq:2.3}), and (\ref{eq:2.5}) are determined solely by $\varphi_{\pi}(u,%
\mu_F^2)$ and the new parameter $\delta^2(\mu_0^2)$. This parameter is
related to the matrix element of the local operator 
\[
\left\langle 0\left| \overline{d}\gamma _{\nu }ig\widetilde{G}_{\mu \rho
}u\right| \pi ^{+}(p)\right\rangle =\frac{1}{3}f_{\pi }\delta ^{2}\left[
p_{\rho }g_{\mu \nu }-p_{\mu }g_{\rho \nu }\right], 
\]
\begin{equation}
\delta ^{2}(\mu _{0}^{2})\simeq 0.2\,\,\mathrm{{GeV}^{2}}  \label{eq:2.8}
\end{equation}
and estimated from the 2-point QCD sum rules \cite{NS94}.

The last problem to be addressed here is a proper choice of the leading
twist DA $\varphi _{\pi }(u,\,\mu _{F}^{2})$. The renormalon method does not
provide a prescription for that case, and we adopt a usual model for $%
\varphi _{\pi }(u,\,\mu _{F}^{2})$ given by a truncated conformal expansion 
\begin{equation}
\varphi _{\pi }(u,\,\mu _{F}^{2})=\varphi _{asy}(u)\left[ 1+B_{2}(\mu
_{F}^{2})C_{2}^{3/2}(u-\overline{u})+ B_{4}(\mu _{F}^{2})C_{4}^{3/2}(u-%
\overline{u})+\cdots \right],  \label{eq:2.9}
\end{equation}
and containing two nonasymptotic terms. In Eq.\ (\ref{eq:2.9}) $\varphi
_{asy}(u)$ is the pion asymptotic DA 
\[
\varphi _{asy}(u)=6u\overline{u}, 
\]
and $C_{n}^{3/2}(\xi )$ are the Gegenbauer polynomials. The functions $%
B_{n}(\mu _{F}^{2})$ determine the evolution of $\varphi _{\pi }(u,\,\mu
_{F}^{2})$ on the factorization scale $\mu _{F}^{2}$, and at the
next-to-leading order (NLO) are given by the following expressions: 
\[
B_2(\mu_F^2)=b_2(\mu_0^2)E_2(\mu_F^2)+\frac{\alpha_{\mathrm{S}}(\mu_F^2)}{%
4\pi}d_0^2(\mu_F^2), 
\]
\begin{equation}  \label{eq:2.10}
B_4(\mu_F^2)=b_4(\mu_0^2)E_4(\mu_F^2)+\frac{\alpha_{\mathrm{S}}(\mu_F^2)}{%
4\pi}\left[d_0^4(\mu_F^2)+ b_2(\mu_0^2)E_2(\mu_F^2)d_2^4(\mu_F^2)\right],
\end{equation}
where $b_2(\mu_0^2)$ and $b_4(\mu_0^2)$ are the input parameters, which
should be extracted from experimental data.

In Eq. (\ref{eq:2.10}) $E_n(\mu_F^2)$ and $d_n^k(\mu_F^2)$ (see Ref.\ \cite%
{DRS}) are defined as

\begin{equation}  \label{eq:2.11}
E_n(\mu_F^2)=\left[ \frac{\alpha_{\mathrm{S}}(\mu _{F}^{2})}{\alpha _{%
\mathrm{S}}(\mu _{0}^{2})}\right] ^{\gamma _{n}/2\beta _{0}}\left[ \frac{%
\beta_0+\beta_1\alpha_{\mathrm{S}}(\mu_F^2)/4\pi} {\beta_0+\beta_1\alpha_{%
\mathrm{S}}(\mu_0^2)/4\pi} \right] ^{(\gamma_1^n /\beta_1-\gamma_0^n/\beta_0
)/2},
\end{equation}
and 
\begin{equation}  \label{eq:2.12}
d_n^k(\mu_F^2)=\frac{M_{nk}}{\gamma_0^k-\gamma_0^n-2\beta_0}\left\{ 1-\left[ 
\frac{\alpha_{\mathrm{S}}(\mu _{F}^{2})}{\alpha _{\mathrm{S}}(\mu _{0}^{2})}%
\right] ^{(\gamma_0^k-\gamma_0^n-2\beta_0)/2\beta _{0}} \right\}.
\end{equation}
Here $\beta_0,\;\gamma_0^n$ and $\beta_1,\;\gamma_1^n$ are the beta function
and the anomalous dimensions one-- and two-loop coefficients, respectively: 
\[
\beta_0=11-\frac{2}{3}n_f,\;\;\beta_1=102-\frac{38}{3}n_f, 
\]
\[
\gamma_0^0=0,\;\gamma_0^2=\frac{100}{9},\;\gamma_0^4=\frac{728}{45}, 
\]
\begin{equation}  \label{eq:2.13}
\gamma_1^0=0,\;\gamma_1^2=\frac{34450}{243}-\frac{830}{81}n_f,\;\gamma_1^4=%
\frac{662846}{3375}- \frac{31132}{2025}n_f,
\end{equation}
with $n_f$ being a number of active quark flavors. The functions $%
d_n^k(\mu_F^2)$ appear in NLO evolution formulas (\ref{eq:2.10}) due to
mixing of partial waves in $\varphi _{\pi }(u,\,\mu _{F}^{2})$ corresponding
to different conformal spins. The numerical values of the matrix $M_{nk}$, 
\[
M_{02}=-11.2+1.73n_f,\;M_{04}=-1.41+0.565n_f,\; M_{24}=-22.0+1.65n_f, 
\]
and the standard two-loop expression for the QCD coupling, 
\begin{equation}  \label{eq:2.14}
\alpha_{\mathrm{S}}(\mu^2)=\frac{4\pi }{\beta_0\ln(\mu^2/\Lambda^2)}\left[1- 
\frac{\beta_1}{\beta_0^2}\frac{\ln \left[\ln(\mu^2/\Lambda^2)\right]} {%
\ln(\mu^2/\Lambda^2)}\right ]
\end{equation}
complete the necessary information on $\varphi _{\pi }(u,\,\mu _{F}^{2})$.

The expansion of $\varphi _{\pi }(u,\,\mu _{F}^{2})$ in conformal spins (\ref%
{eq:2.9}) is the standard prescription in PQCD and is widely used in
applications. However, to calculate the DAs $\varphi_1^{(4)}(u,\mu_F^2)$, $%
\varphi_2^{(4)}(u,\mu_F^2)$, as well as the twist-4 contribution to $%
F_{\pi\gamma}(Q^2)$, we shall use the expansion of $\varphi _{\pi }(u,\,\mu
_{F}^{2})$ in powers of $u$, 
\begin{equation}
\varphi _{\pi }(u,\,\mu _{F}^{2})=\varphi _{asy}(u)\sum_{n=0}^{4 }K_{n}(\mu
_{F}^{2})u^{n}.  \label{eq:2.15}
\end{equation}
The coefficients $K_{n}(\mu _{F}^{2})$ in Eq.\ (\ref{eq:2.15}) are given by
the following equalities: 
\[
K_{0}(\mu _{F}^{2})=1+6B_{2}(\mu _{F}^{2})+15B_{4}(\mu
_{F}^{2}),\,\,K_{1}(\mu _{F}^{2})=-30\left[ B_{2}(\mu _{F}^{2})+7B_{4}(\mu
_{F}^{2})\right], 
\]
\[
K_2(\mu _F^2)=30\left[ B_2(\mu _F^2)+28B_4(\mu _F^2)\right] ,\,\,K_3(\mu
_F^2)=-1260B_4(\mu _F^2),\, 
\]
\begin{equation}
\,K_{4}(\mu _{F}^{2})=630B_{4}(\mu _{F}^{2}).  \label{eq:2.16}
\end{equation}

The distributions $\varphi_1^{(4)}(u,\mu_F^2)$ and $\varphi_2^{(4)}(u,%
\mu_F^2)$ in the renormalon approach were found in our work \cite{A02} using
the expansion (\ref{eq:2.15}) \footnote{%
The term $-61u^3\overline{u}^3/45$ in $\varphi_4^1(u)$ (Ref. \cite{A02}, Eq.
(2.18)) should be replaced by $-41u^3\overline{u}^3/36$. But this correction
does not affect results and conclusions of Ref. \cite{A02}.} 
\[
\varphi _{1}^{(4)}(u,\mu _{F}^{2})=\sum_{n=0}^{4}K_{n}(\mu _{F}^{2})\varphi
_{n}^{1}(u), 
\]
\begin{equation}
\varphi _{2}^{(4)}(u,\mu _{F}^{2})=\sum_{n=0}^{4}K_{n}(\mu _{F}^{2})\varphi
_{n}^{2}(u).  \label{eq:2.17}
\end{equation}%
The explicit expressions of their components $\varphi_n^1(u)$ and $%
\varphi_n^2(u)$ are collected in the Appendix.

\section{The form factor $F_{\protect\pi\protect\gamma}(Q^2)$ in the QCD
LCSR method}

For the calculation of the electromagnetic $\gamma\gamma^*\to \pi^0 $ form
factor $F_{\pi\gamma}(Q^2)$ in the present work, we use the QCD LCSR method,
which is one of the effective tools to estimate nonperturbative components
of exclusive quantities \cite{BBK89}. The LCSR expression for the FF $%
F_{\pi\gamma}(Q^2)$ was derived in Ref.\ \cite{Kd99}, where its tree-level
twist-2 and twist-4 components were found. The $O(\alpha_{\mathrm{S}})$
correction to the twist-2 part was computed in Ref.\ \cite{SY00}, and on the
basis of these results constraints on the parameters $b_2(\mu_0^2)$ and $%
b_4(\mu_0^2)$ were extracted from CLEO data. This analysis was refined
recently in Ref.\ \cite{BMS03}, where a new model for the pion leading twist
DA was proposed. The renormalon approach to the twist-4 term leads to
further insight on the FF, because it allows one to take into account
effects due to the higher conformal spins neglected in the previous studies.

The LCSR method is based on the analysis of the correlation function of the
transition $\gamma^*(q_1)\gamma^*(q_2)\to \pi^0(p)$ \cite{Kd99} 
\begin{equation}  \label{eq:3.1}
\int d^4xe^{-iq_1x}\left\langle \pi^0(p)\left| T\left\{ j_\mu(x)j_\nu
(0)\right\} \right| 0\right\rangle=
i\epsilon_{\mu\nu\alpha\beta}q_1^{\alpha}q_2^{\beta}F_{\pi\gamma^*}(Q^2,q^2),
\end{equation}
where $Q^2=-q_1^2,\;q^2=-q_2^2$ are the virtualities of the photons, $%
j_{\mu}=e_u \overline{u}\gamma _\mu u+e_d\overline{d}\gamma _\mu d$ is the
quark electromagnetic current and $F_{\pi\gamma^*}(Q^2,q^2)$ is the form
factor of the transition $\gamma^*(q_1)\gamma^*(q_2)\to \pi^0(p)$.

For large values of $Q^2$ and $q^2$, this correlator can be computed in
PQCD. In the QCD sum rules method by matching between the dispersion
relation in terms of contributions of hadronic states, which include a
contribution of the low-lying physical states in the $q$-channel, i.e., that
due to the vector $\rho$ and $\omega$ mesons, as well as a contribution
coming from the continuum of hadronic states with the same quantum numbers,
and the QCD calculation at Euclidean momenta, one can estimate the form
factor $F_{\pi\gamma^*}(Q^2,q^2)$.

After calculations, in the limit $q^2 \to0$ the formula for the FF $%
F_{\pi\gamma}(Q^2)$ can be obtained: 
\begin{equation}  \label{eq:3.2}
F_{\pi\gamma}(Q^2)=\frac{1}{\pi m_{\rho}^2}\int_0^{s_0}ds\ \mathrm{Im}%
F_{\pi\gamma^*}(Q^2,s) \exp \left(\frac{m_{\rho}^2-s}{M^2}\right ) +\frac{1}{%
\pi}\int_{s_0}^{\infty}\frac{ds}{s}\ \mathrm{Im}F_{\pi\gamma^*}(Q^2,s),
\end{equation}
where\footnote{
Starting from Eq.\ (\ref{eq:3.3}) $\mu_F^2=Q^2$.} 
\begin{equation}  
\label{eq:3.3}
\frac{1}{\pi}\mathrm{Im}F_{\pi\gamma^*}(Q^2,s)=\frac{\sqrt{2}f_{\pi}}{3}%
\left [ \frac{\varphi^{(2)}(u,Q^2)}{s+Q^2} -\frac{1}{Q^2}\frac{%
d\Phi^{(4)}(u,Q^2)}{ds} \right ]_{u=\frac{Q^2}{s+Q^2}}.
\end{equation}
In Eq.\ (\ref{eq:3.2}) $M^2$ is the Borel parameter, $m_{\rho}$ is the mass
of the $\rho$-meson (and $p^2=m_\pi^2=0$), and $s_0$ is the duality
interval. By $\Phi^{(4)}(u,Q^2)$ the following combination of the twist-4
DAs is denoted: 
\[
\Phi^{(4)}(u,Q^2)=4(\varphi_1^{(4)}(u,Q^2)-\varphi_2^{(4)}(u,Q^2)) 
\]
\begin{equation}  \label{eq:3.4}
+\int_0^ud\alpha_1\int_0^{1-u}\frac{d\alpha_2}{\alpha_3} \left [\frac{%
1-2u+\alpha_1-\alpha_2}{\alpha_3}\Phi_{\|}(\alpha_1,\alpha_2,\alpha_3)-%
\Psi_{\|}(\alpha_1,\alpha_2,\alpha_3) \right]_{\alpha_3=1-\alpha_1-%
\alpha_2}.
\end{equation}

The first term in the right-hand side of Eq.\ (\ref{eq:3.3}) gives rise to
the tree-level twist-2 component $F_{\pi\gamma}^{(2)}(Q^2)$ of the FF,
whereas the second one generates the twist-4 contribution $%
F_{\pi\gamma}^{(4)}(Q^2)$. In general, the $F_{\pi\gamma} (Q^2)$ has the
following form: 
\begin{equation}  \label{eq:3.5}
F_{\pi\gamma} (Q^2)=F_{\pi\gamma} ^{(2)}(Q^2)+F_{\pi\gamma} ^{(2,\alpha _{%
\mathrm{S}})}(Q^2)+F_{\pi\gamma} ^{(4)}(Q^2),
\end{equation}
where $F_{\pi\gamma} ^{(2,\alpha _{\mathrm{S}})}(Q^2)$ is the $O(\alpha_{%
\mathrm{S}})$ correction to the twist-2 term. The formulas that determine $%
F_{\pi\gamma}^{(2,\alpha \mathrm{{_S})}}(Q^2)$ are cumbersome and not
written down here. Their explicit expressions can be found in Refs.\ \cite%
{SY00,BMS03}.

The twist-2 and --4 components of the FF can be rewritten in the form
convenient for further analysis \cite{Kd99}: 
\begin{equation}  \label{eq:3.6}
Q^2F_{\pi\gamma}^{(2)}(Q^2)=\frac{\sqrt{2}f_\pi}{3}\left\{\frac{Q^2}{m_\rho^2%
}\int_{u_0}^1\frac{du}{u} \varphi^{(2)}(u,Q^2)\exp\left[-\frac{Q^2\overline{u%
}}{uM^2}+\frac{m_\rho^2}{M^2}\right] +\int_0^{u_0}\frac{du}{\overline{u}}%
\varphi^{(2)}(u,Q^2) \right\},
\end{equation}
and 
\begin{equation}  \label{eq:3.7}
Q^2F_{\pi\gamma}^{(4)}(Q^2)=\frac{\sqrt{2}f_\pi}{3}\left\{\frac{1}{m_\rho^2}%
\int_{u_0}^1du \frac{d\Phi^{(4)}(u,Q^2)}{du}\exp\left[-\frac{Q^2\overline{u}%
}{uM^2}+\frac{m_\rho^2}{M^2}\right] +\frac{1}{Q^2}\int_0^{u_0}du\frac{u}{%
\overline{u}}\frac{d\Phi^{(4)}(u,Q^2)}{du} \right\},
\end{equation}
where $u_0=Q^2/(s_0+Q^2)$.

Our main interest is the twist-4 term and its calculation using the
renormalon-inspired twist-4 DAs. The DAs entering to Eq.\ (\ref{eq:3.4}) are
known and after some calculations we get 
\begin{equation}  \label{eq:3.8}
\Phi^{(4)}(u,Q^2)=\sum_{n=0}^4K_n(Q^2)\Phi_n^{(4)}(u),
\end{equation}
where the components $\Phi_n^{(4)}(u)$ are 
\[
\Phi_0^{(4)}(u)=2\delta^2\left[-2u^2 \ln u-2\overline{u}^2\ln \overline{u}%
\right],\;\; \Phi_1^{(4)}(u)=2\delta^2\frac{2}{3}\left[u\overline{u}-2u^3
\ln u- 2\overline{u}^3\ln \overline{u}\right], 
\]
\[
\Phi_2^{(4)}(u)=2\delta^2 \left[u\overline{u}\left (1-\frac{1}{6}u\overline{u%
}\right) +u^2(1-2u)\ln u +\overline{u}^2(1-2\overline{u})\ln \overline{u} %
\right], 
\]
\[
\Phi_3^{(4)}(u)=2\delta^2 \left\{\frac{u\overline{u}}{30}\left[36-u\overline{%
u}-10u^2\overline{u}^2 \right] +2u^2\left[1-2u+u^2-\frac{2}{5}u^3 \right]\ln
u+ 2\overline{u}^2\left[1-2\overline{u}+\overline{u}^2-\frac{2}{5}\overline{u%
}^3 \right]\ln \overline{u} \right\}, 
\]
\begin{equation}  \label{eq:3.9}
\Phi_4^{(4)}(u)=2\delta^2 \left\{u\overline{u}\left[\frac{4}{3}+\frac{7}{20}u%
\overline{u}- \frac{107}{90}u^2\overline{u}^2 \right] +u^2\left[3-\frac{20}{3%
}u+5u^2-2u^3 \right]\ln u+ \overline{u}^2\left[3-\frac{20}{3}\overline{u}+5%
\overline{u}^2-2\overline{u}^3 \right] \ln \overline{u} \right\}.
\end{equation}
The twist-4 function (\ref{eq:3.8}) coincides with one obtained in Ref. \cite%
{BMS05}.

%%%%%%%%%%%%%%%%%%%%%%%%%%%%%%%%%%%%%%%%%%%%%%%%%%%%%%%%%%%%%%%%%%%%%
%                            F I G U R E  1                         %
%                          \label{fig:rgfig1}                        %
%%%%%%%%%%%%%%%%%%%%%%%%%%%%%%%%%%%%%%%%%%%%%%%%%%%%%%%%%%%%%%%%%%%%%
\begin{figure}[t]
%%Fig 1
\centering\epsfig{file=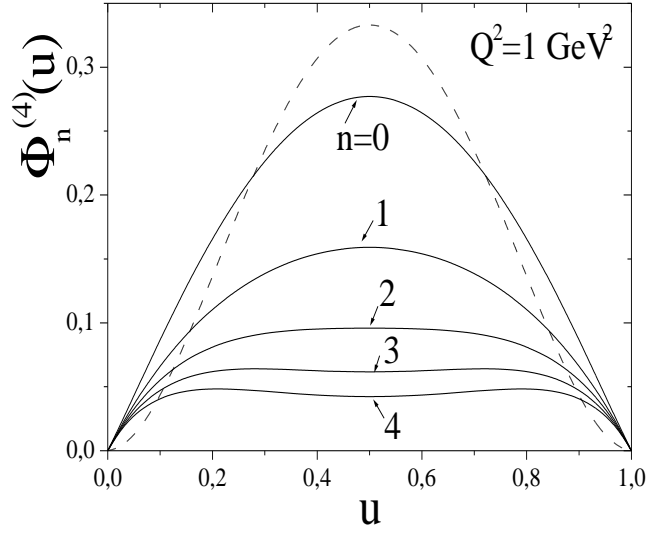,height=7cm,width=8.5cm,clip=}
\caption{ The components $\Phi_n^{(4)}(u)$ of $\Phi^{(4)}(u,Q^2)$ (\ref%
{eq:3.8}) as functions of $u$. For comparison, the asymptotic DA (\ref%
{eq:3.10}) is also shown (the dashed line). }
\label{fig:rgfig1}
\end{figure}
%%%%%%%%%%%%%%%%%%%%%%%%%%%%%%%%%%%%%%%%%%%%%%%%%%%%%%%%%%%%%%%%%%%%%
In Refs.\ \cite{Kd99,SY00,BMS03} the twist-4 contribution to $%
F_{\pi\gamma}(Q^2)$ was estimated using the asymptotic form of the relevant
twist-4 DAs, which lead to the simple expression 
\begin{equation}  \label{eq:3.10}
\Phi_{asy}^{(4)}(u,Q^2)=\frac{80}{3}\delta^2(Q^2)u^2\overline{u}^2,\;\;\;\;
\delta^2(Q^2)=\delta^2(\mu_0^2)\left[\frac{\alpha_{\mathrm{S}}(Q^2)}{\alpha_{%
\mathrm{S}}(\mu_0^2)} \right]^{8C_F/3\beta_0}.
\end{equation}
The function $\Phi_{asy}^{(4)}(u,Q^2)$ and components of $\Phi^{(4)}(u,Q^2)$
are shown in Fig.\ \ref{fig:rgfig1}. Without any detailed analysis, the
difference in their behavior in the end-point regions $u=0;1$ is evident: we
have emphasized that the renormalon model predicts DAs that are larger at
the end points. The difference between them becomes more essential, when
considering the behavior of $d\Phi^{(4)}(u,Q^2)/du$ in the end-point
regions. Thus, for the asymptotic model $d\Phi_{asy}^{(4)}(u,Q^2)/du=0$ at $%
u=0;1$. In the case of the renormalon-inspired DAs for $n=0-4$ we find $%
d\Phi_n^{(4)}(u)/du=4\delta^2(Q^2)$ at $u=0$ and $d\Phi_n^{(4)}(u)/du=-4%
\delta^2(Q^2)$ at $u=1$.

The expression of the twist-4 contribution (\ref{eq:3.7}) through the
spectral density $\rho^{(4)}(Q^2,s)$ and that of $\rho^{(4)}(Q^2,s)$ itself
are presented in the Appendix.

\section{Extracting the pion DAs from the expeimental data}

The LCSR expression for the pion electromagnetic transition FF can be used
to extract constraints on the input parameters $b_{2}(\mu _{0}^{2})$ and $%
b_{4}(\mu _{0}^{2})$ of the leading twist DA. In order to perform numerical
computations, we fix various parameters appearing in the relevant
expressions. Namely, we take the Borel parameter $M^{2}$ within the interval 
$0.7\,\, \mathrm{{GeV}^2 <M^{2}<1.4\,\,{GeV}^{2}}$ and for the factorization
and renormalization scales accept 
\[
\mu _{F}^{2}=\mu _{R}^{2}=Q^{2}. 
\]
For the QCD coupling $\alpha_{\mathrm{S}}(Q^{2})$ the two-loop expression (%
\ref{eq:2.14}) with $\Lambda _{3}=0.34\,\,\mathrm{GeV}$ is used. The duality
parameter $s_{0}=1.5\,\,\mathrm{{GeV}^{2}}$ is determined from the two-point
sum rules in the $\rho$-meson channel \cite{SVZ}. The normalization scale is
set equal to $\mu_0^2=1 \; \mathrm{{GeV}^2}$. We also use $m_{\rho}=0.775\ 
\mathrm{GeV}$ and $f_{\pi}=0.132\ \mathrm{GeV}$.

%%%%%%%%%%%%%%%%%%%%%%%%%%%%%%%%%%%%%%%%%%%%%%%%%%%%%%%%%%%%%%%%%%%%%
%                            F I G U R E  2                         %
%                          \label{fig:rgfig2}                        %
%%%%%%%%%%%%%%%%%%%%%%%%%%%%%%%%%%%%%%%%%%%%%%%%%%%%%%%%%%%%%%%%%%%%%
\begin{figure}[t]
%%Fig 2
\centering\epsfig{file=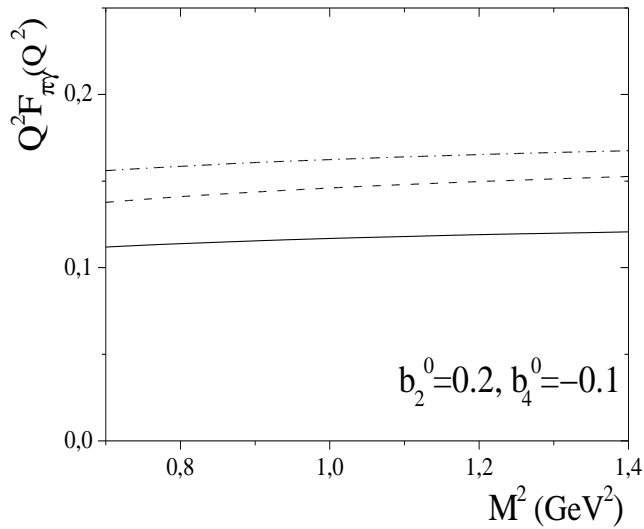,height=7cm,width=8.5cm,clip=}
\caption{ The dependence of the scaled FF $Q^2F_{\protect\pi\protect\gamma%
}(Q^2)$ on the Borel parameter $M^2$. The DA with $b_2(\protect\mu%
_0^2)=0.2,\;b_4(\protect\mu_0^2)=-0.1$ is used. For the solid curve $%
Q^{2}=2\,\mathrm{{GeV}^{2}}$, for the dashed curve $Q^{2}=4\,\mathrm{{GeV}%
^{2}}$, and for the dot-dashed one $Q^{2}=8\,\mathrm{{GeV}^{2}}$.}
\label{fig:rgfig2}
\end{figure}
%%%%%%%%%%%%%%%%%%%%%%%%%%%%%%%%%%%%%%%%%%%%%%%%%%%%%%%%%%%%%%%%%%%%%

The Borel parameter dependence of the LCSR for different values of $Q^2$ is
depicted in Fig.\ \ref{fig:rgfig2}. In calculations the leading twist DA
with $b_2(\mu_0^2)\equiv b_2^0=0.2,\;b_4(\mu_0^2)\equiv b_4^0=-0.1$, as well
as the twist-4 function (\ref{eq:3.8}) are used. From this figure, one can
conclude that the prediction for the FF is rather stable in the exploring
range of $M^2$. In what follows we choose the Borel parameter equal to $%
M^2=1\,\,\mathrm{{GeV}^2}$.

%%%%%%%%%%%%%%%%%%%%%%%%%%%%%%%%%%%%%%%%%%%%%%%%%%%%%%%%%%%%%%%%%%%%%
%                            F I G U R E 3                         %
%                          \label{fig:rgfig3}                        %
%%%%%%%%%%%%%%%%%%%%%%%%%%%%%%%%%%%%%%%%%%%%%%%%%%%%%%%%%%%%%%%%%%%%%
\begin{figure}[tb]
%%Fig 3
\centering\epsfig{file=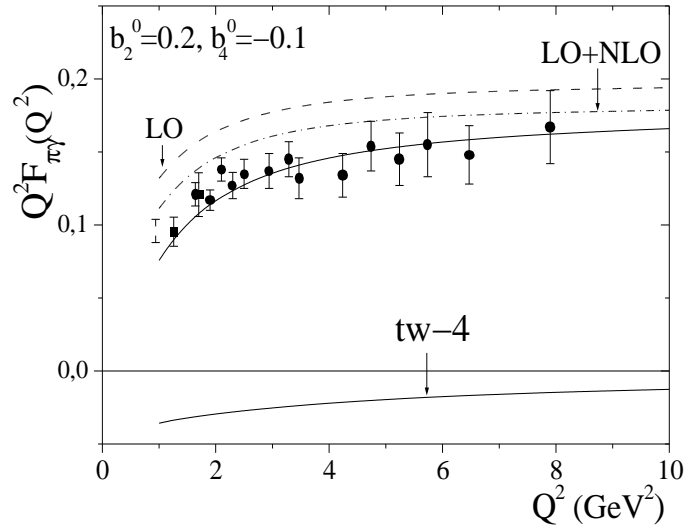,height=7cm,width=9cm,clip=}
\caption{ The scaled transition FF as a function of $Q^2$. The solid line
corresponds to the sum of all the contributions (\ref{eq:3.5}). The data are
taken from Refs.\ \protect\cite{CELLO} (the rectangles) and \protect\cite%
{CLEO} (the circles). }
\label{fig:rgfig3}
\end{figure}
%%%%%%%%%%%%%%%%%%%%%%%%%%%%%%%%%%%%%%%%%%%%%%%%%%%%%%%%%%%%%%%%%%%%%
The scaled FF $Q^{2}F_{\pi\gamma }(Q^{2})$ and its different components are
plotted in Fig.\ \ref{fig:rgfig3}. As is seen, the leading order prediction
for the FF in the QCD LCSR method considerably overshoots the data points.
By including into consideration the NLO correction, one may only soften this
discrepancy, but not remove it. In fact, the LCSR prediction with LO+NLO
accuracy, for the values of the input parameters $b_2^0,\;b_4^0$ shown in
the figure, again overestimates the experimental data. It is evident that
DAs with $b_4^0\geq 0$ will lead to a more great deviation from the
experimental data than DAs with $b_4^0<0$. Therefore, the traditional
treatment of the transition FF \cite{SY00,BMS03} would call for DAs with $%
b_4^0<0$, because the asymptotic twist-4 contribution is not strong enough
to compensate the growth of the LO+NLO term if $b_4^0 \geq 0$. 
In the renormalon approach the
situation, in general, remains the same. This fact is connected with the
dependence of the twist-4 term on the input parameters $b_2^0,\;b_4^0$.

\begin{figure}[tb]
%%Fig 4
\centering\epsfig{file=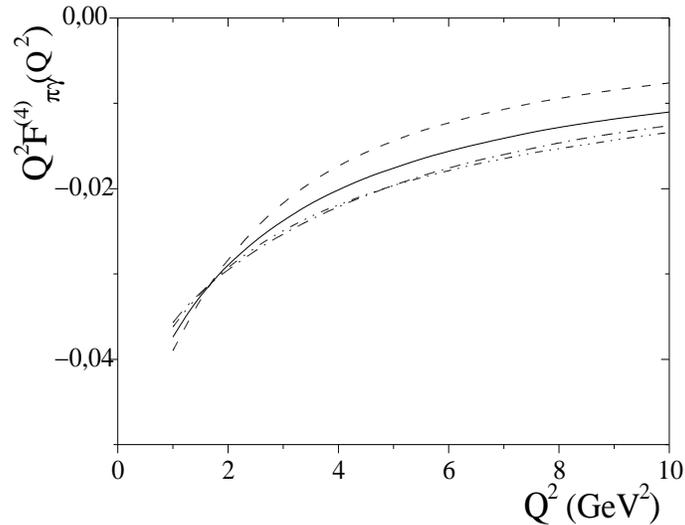,height=7cm,width=9cm,clip=}
\caption{ The twist-4 term as a function of $Q^{2}$. The dashed curve is
computed using the function $\Phi_{asy}^{(4)}(u,Q^2)$. The predictions for
the twist-4 term obtained employing the renormalon-inspired DAs are shown by
the solid and two broken lines. The correspondence between the lines and the
input parameters is: the solid line, $b_{2}(\protect\mu_{0}^{2})=0,\,\,b_{4}(%
\protect\mu _{0}^{2})=0$; the dot-dashed line, $b_{2}(\protect\mu%
_{0}^{2})=0.2,\,\,b_{4}(\protect\mu _{0}^{2})=-0.1$; the dot-dot-dashed
line, $b_{2}(\protect\mu _{0}^{2})=0.2,\,\,b_{4}(\protect\mu _{0}^{2})=0.1$.}
\label{fig:rgfig4}
\end{figure}

To understand this important point, it is instructive to explore the twist-4
contributions corresponding to different DAs. The relevant results are shown
in Fig.\ \ref{fig:rgfig4}. Here the dashed and solid curves are computed
using the standard asymptotic function (\ref{eq:3.10}) and the
renormalon-generated one (\ref{eq:3.8}) with $b_{2}^{0}=b_{4}^{0}=0,$
respectively. The difference between them is evident: almost in the whole
region of the explored momentum transfers $Q^{2}\geq 1.8\;\mathrm{{GeV}^{2}}$
the higher conformal spin (renormalon) effects enhance the absolute value of
the twist-4 contribution. The twist-4 terms corresponding to $b_{2}^{0}\neq
0,\;b_{4}^{0}\neq 0$ (the dot-dashed and dot-dot-dashed curves) in the
region $1\ \mathrm{{GeV}^{2}\leq Q^{2}\leq 1.8\ {GeV}^{2}}$ are larger than
the asymptotic contribution, whereas for $Q^{2}>1.8\ \mathrm{{GeV}^{2}}$
they run below both the solid and dashed curves. It is worth noting that at
fixed $b_{2}^{0}=0.2$, curves $b_{4}^{0}=\pm 0.1$ (these values have been
chosen as sample ones ) are close to each other: the sizeable difference
between them appears only at $Q^{2}\geq 8\ \mathrm{{GeV}^{2}}$. This means
that in the low-$Q^{2}$ region, DAs with $b_{4}^{0}=\pm 0.1$ reduce the
LO+NLO contribution almost in the same manner, and this effect is smaller
than the corresponding effect in the case of the asymptotic contribution. In
the domain of the high $Q^{2}$, the twist-4 terms cut the LO+NLO
result more effectively than the asymptotic twist-4 term, but because 
the LO+NLO ($b_{4}^{0}<0$) contribution  itself is smaller 
than the LO+NLO ($b_{4}^{0}\geq 0$) one, it terns out that only DAs with $%
b_{4}^{0}<0$ lead to agreement with the data.

%%%%%%%%%%%%%%%%%%%%%%%%%%%%%%%%%%%%%%%%%%%%%%%%%%%%%%%%%%%%%%%%%%%%%
%                            F I G U R E 4                         %
%                          \label{fig:rgfig4}                        %
%%%%%%%%%%%%%%%%%%%%%%%%%%%%%%%%%%%%%%%%%%%%%%%%%%%%%%%%%%%%%%%%%%%%%
%%%%%%%%%%%%%%%%%%%%%%%%%%%%%%%%%%%%%%%%%%%%%%%%%%%%%%%%%%%%%%%%%%%%%

\begin{figure}[tb]
%%Fig 5
\centering\epsfig{file=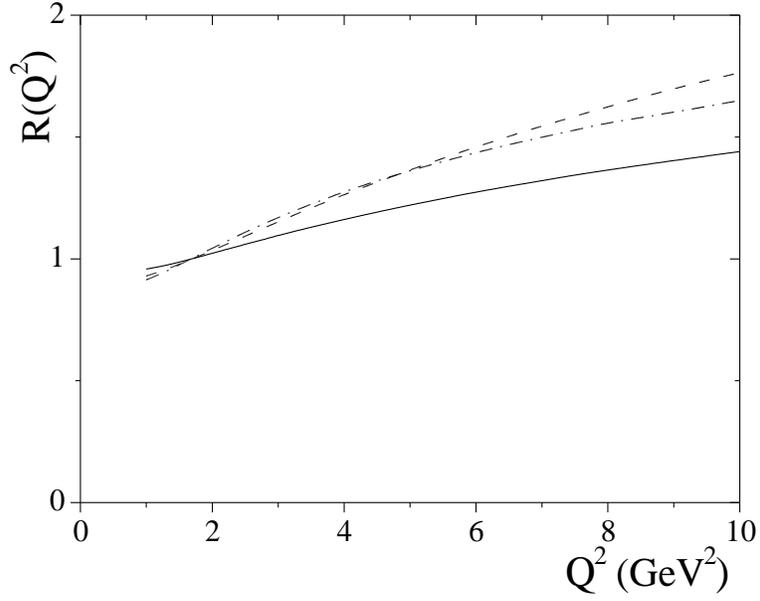,height=8cm,width=10cm,clip=}
\caption{ The ratio $R(Q^2)$ as a function of $Q^2$. The solid line
corresponds to the input parameters $b_2(\protect\mu_0^2)=b_4(\protect\mu%
_0^2)=0$. The dashed line describes the same ratio, but for $b_2(\protect\mu%
_0^2)=0.2,\;b_4(\protect\mu_0^2)=0.1$, while the dot-dashed one corresponds
to $b_2(\protect\mu_0^2)=0.2,\;b_4(\protect\mu_0^2)=-0.1$.}
\label{fig:rgfig5}
\end{figure}

In general, calculations of Ref.\ \cite{SY00,BMS03} correspond essentially
to the ''minimal'' model of the twist-4 effects, where the restriction to
the lowest conformal spin probably underestimates the effect, while the
renormalon model is a ''maximal'' model, where these effects are probably
somewhat overestimated. Therefore, the renormalon model for the twist-4 DAs
allows us to put a theoretically justified bound on the twist-4 contribution
to the pion transition form factor. The change in absolute value of the
twist-4 term is not dramatic, as it may be expected. To quantify this
statement we introduce the ratio $R(Q^2)$, 
\[
R(Q^2)=\frac{[F_{\pi\gamma}^{(4)}(Q^2)]^{\mathrm{ren}}}{[F_{\pi%
\gamma}^{(4)}(Q^2)]^{\mathrm{stand}}}, 
\]
and demonstrate its numerical results in Fig.\ \ref{fig:rgfig5} for some
selected values of $b_2^0, \; b_4^0$.

%%%%%%%%%%%%%%%%%%%%%%%%%%%%%%%%%%%%%%%%%%%%%%%%%%%%%%%%%%%%%%%%%%%%%
%                            F I G U R E 5                         %
%                          \label{fig:rgfig5}                        %
%%%%%%%%%%%%%%%%%%%%%%%%%%%%%%%%%%%%%%%%%%%%%%%%%%%%%%%%%%%%%%%%%%%%%
%%%%%%%%%%%%%%%%%%%%%%%%%%%%%%%%%%%%%%%%%%%%%%%%%%%%%%%%%%%%%%%%%%%%%

%%%%%%%%%%%%%%%%%%%%%%%%%%%%%%%%%%%%%%%%%%%%%%%%%%%%%%%%%%%%%%%%%%%%%
%                            F I G U R E 6                         %
%                          \label{fig:rgfig6}                        %
%%%%%%%%%%%%%%%%%%%%%%%%%%%%%%%%%%%%%%%%%%%%%%%%%%%%%%%%%%%%%%%%%%%%%
\begin{figure}[b]
%%Fig 6
\centering\epsfig{file=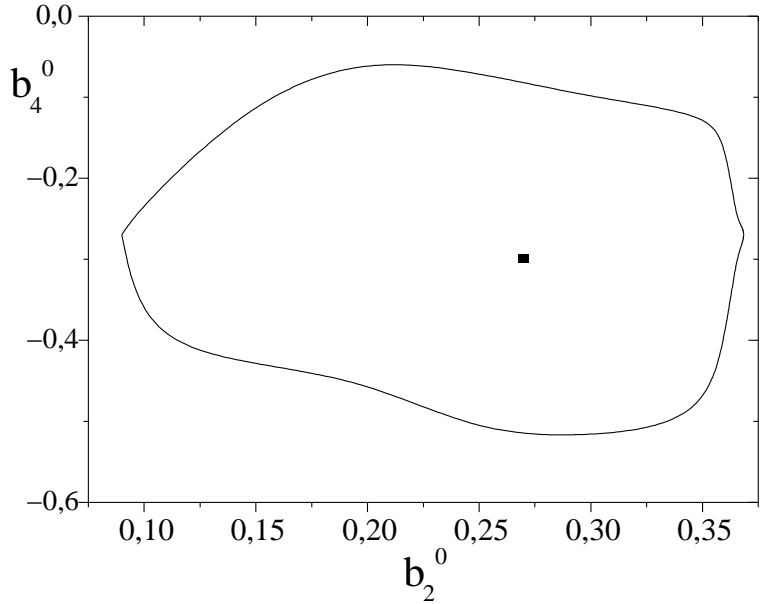,height=8cm,width=10cm,clip=}
\caption{ The $1\protect\sigma $ area in the $b_{2}^{0}-b_{4}^{0}$ plane
extracted from comparison of the CELLO and CLEO data with the LCSR
prediction. The central solid rectangle denotes the point $%
b_{2}^{0}=0.27,\;b_{4}^{0}=-0.3$.}
\label{fig:rgfig6}
\end{figure}
%%%%%%%%%%%%%%%%%%%%%%%%%%%%%%%%%%%%%%%%%%%%%%%%%%%%%%%%%%%%%%%%%%%%%
In Fig.\ \ref{fig:rgfig6} we plot the $1\sigma $ area in the $%
b_{2}^{0}-b_{4}^{0}$ plane, extracted at the scale $\mu _{0}^{2}=1\ \mathrm{{%
GeV}^{2}}$ in the result of the fitting procedure. One can see that this
area stretches in the lower-half plane occupying a large region. The
parameter $b_{2}^{0}$ varies within the limits%
\[
b_{2}^{0}\in \lbrack 0.09,\,\,0.37], 
\]%
whereas at some fixed $b_{2}^{0}$ the variation of $b_{4}^{0}$ is
even larger. For example, at $b_{2}^{0}=0.2$, $b_{4}^{0}$ takes values in
the region%
\[
b_{2}^{0}\in \lbrack -0.05,-\,\,0.45]. 
\]

The $1\sigma $ region for the transition FF is shown in Fig.\ \ref%
{fig:rgfig7}. In the fitting procedure we employ the CLEO data, and the two
CELLO data points at $Q^{2}=1.26$ and $1.7\ \mathrm{{GeV}^{2}}$. The CELLO
data have an important impact on the fitting procedure.

The central values for the parameters $b_{2}^{0}$ and $b_{4}^{0}$ extracted
in this work are 
\begin{equation}
b_{2}^{0}(1\ \mathrm{GeV}^{2})=0.27,\,\,b_{4}^{0}(1\ \mathrm{{GeV}^{2})=-0.3.%
}  \label{eq:4.1}
\end{equation}

\bigskip It is difficult to present the $1\sigma $ region of the parameters
by a simple formula, therefore we refrain from such attemps.

\begin{figure}[tb]
%%Fig 7
\centering\epsfig{file=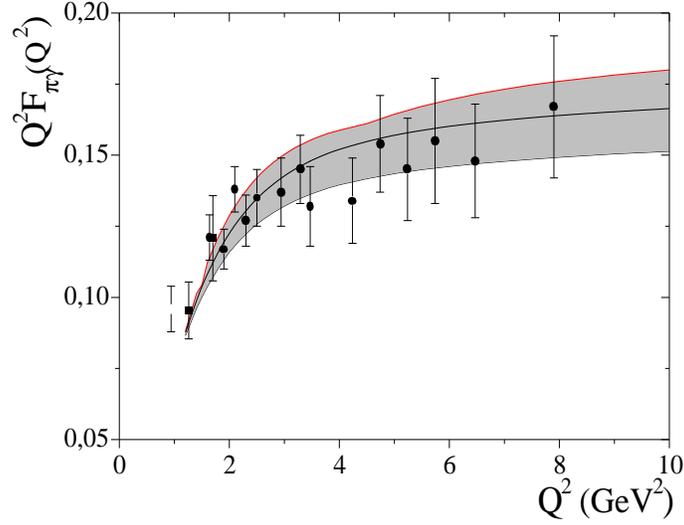,height=7cm,width=9cm,clip=}
\caption{ The form factor $Q^{2}F_{\protect\pi \protect\gamma }(Q^{2})$ as a
function of $Q^{2}$. The shaded area demonstrates $1\protect\sigma $ region
for the transition FF. In the fitting procedure the solid data points have
been used. The central solid curve corresponds to the parameters $%
b_{2}^{0}=0.27,\;b_{4}^{0}=-0.3$. }
\label{fig:rgfig7}
\end{figure}

As is seen the curve in Fig. \ref{fig:rgfig7}  in the low-$Q^{2}$ region
decreases rapidly, deviating considerably from the CELLO point 
at $Q^{2}=0.94\; \mathrm{{GeV}^{2}}$: 
the DA (\ref{eq:4.1}) \ describes the CELLO and CLEO\
data located only in the domain $1.26\;\mathrm{GeV}^{2}\leq Q^{2}\leq 7.9\;%
\mathrm{GeV}^{2}$. 

Evolved to the normalization scale $\mu _{0}^{2}=5.76\ \mathrm{{GeV}^{2}}$,
the parameters (\ref{eq:4.1}) become equal to 
\begin{equation}
b_{2}^{0}\simeq 0.184,\,\,b_{4}^{0}\simeq -0.179.  \label{eq:4.2}
\end{equation}%
Equation (\ref{eq:4.2}) can be compared with the predictions ($\mu
_{0}^{2}=5.76\ \mathrm{GeV}^{2}$) from Ref.\ \cite{SY00}, 
\begin{equation}
b_{2}^{0}=0.19,\,\,b_{4}^{0}=-0.14,  \label{eq:4.2a}
\end{equation}%
and Ref.\ \cite{BMS03} 
\begin{equation}
b_{2}^{0}=0.2,\,\,b_{4}^{0}=-0.17.  \label{eq:4.2b}
\end{equation}%
From this analysis it becomes evident that both the standard and renormalon
approaches predict the pion leading twist DA with a negative coefficient $%
b_{4}^{0}<0.$

%%%%%%%%%%%%%%%%%%%%%%%%%%%%%%%%%%%%%%%%%%%%%%%%%%%%%%%%%%%%%%%%%%%%%
%                            F I G U R E 7                         %
%                          \label{fig:rgfig7}                        %
%%%%%%%%%%%%%%%%%%%%%%%%%%%%%%%%%%%%%%%%%%%%%%%%%%%%%%%%%%%%%%%%%%%%%
%%%%%%%%%%%%%%%%%%%%%%%%%%%%%%%%%%%%%%%%%%%%%%%%%%%%%%%%%%%%%%%%%%%%%
In our previous paper \cite{A02} we obtained the constraints on the pion
leading twist DA from LCSR analysis of the pion electromagnetic FF. In
calculations we employed the renormalon model for the twist-4 distributions (%
\ref{eq:2.17}), but took into account the evolution of the pion DAs with the
lower than in the present work accuracy (i.e., with the LO accuracy).  At
the scale $\mu _{0}^{2}=1\ \mathrm{{GeV}^{2}}$ we found 
\begin{equation}
b_{2}^{0}=0.2\pm 0.03,\,\,b_{4}^{0}=-0.03\pm 0.06.  \label{eq:4.3}
\end{equation}%
The $1\sigma $ area depicted in Fig.\ \ref{fig:rgfig6} overlaps with \ (\ref%
{eq:4.3}) (it is not shown in the figure) in the region determined by the
following values of the parameters   
\[
b_{2}^{0}\simeq 0.2-0.23,\;b_{4}^{0}\simeq -0.05-(-0.09).
\]%
In general, an $1\sigma $ area for two quantities is not necessarily equal
to the overlap region of two independent $1\sigma $ areas. Nevertheless, it
is important, at least as the first approximation, to determine the pion DA
that satisfactorily describes the both of these form factors. In order to
get more precise estimate, $F_{\pi }(Q^{2})$ has to be calculated with
higher accuracy, and a joint treatment of $F_{\pi \gamma }(Q^{2})$ and $%
F_{\pi }(Q^{2})$ must be performed. These tasks are beyond the scope of the
present work.

\section{Conclusions}

In this work we have used the pion renormalon-inspired higher twist DAs to
calculate the twist-4 contribution to the transition form factor $%
F_{\pi\gamma}(Q^2)$. The renormalon method has allowed us to express two--
and three-particle twist-4 DAs in terms of the pion leading twist DA and one
additional parameter $\delta^2$. In other words, in this method the twist-4
distributions are determined by the twist-2 one unambiguously, that
restricts a freedom in the choice of DAs, increasing, at the same time, the
predictive power and reliability of QCD results.

The higher twist distributions introduced in Ref.\ \cite{BGG04} embrace
higher conformal spin effects which so far escaped attention. They are
larger at the end points and, as expected, give rise to larger higher twist
effects in exclusive reactions. Our results on the twist-4 contribution to
the transition FF $F_{\pi \gamma }(Q^{2})$ confirm this assumption. Indeed,
in the region of high momentum transfers the absolute value of the
renormalon-generated twist-4 term exceeds the asymptotic one by a factor $%
1.5-2$. Nevertheless, the LCSR results obtained using the
renormalon-inspired and standard DAs predict the pion leading twist DAs with 
$b_{4}^{0}<0$. Stated differently, effects due to higher conformal spin
components of the twist-4 DAs remain under control and do not spoil the QCD
LCSR method.

The new contribution of this work is that the renormalon approach has
allowed one to put an upper bound on the twist-4 contribution to the
light-cone sum rules and obtain estimates of the effects due to higher
conformal spins.On the example of the $\gamma \gamma^{*} \to \pi^0$ 
transition, it also demonstrated limits of theoretical uncertainties
inherent to our knowledge of exclusive processes.

\begin{acknowledgments}
The author would like to thank Prof.\ V.\ M.\ Braun for illuminating
discussions, valuable comments on the manuscript, and Dr.\ A.\ Manashov for
useful remarks. He also appreciates hospitality of the members of the
Theoretical Physics Institute extended to him in Regensburg, where this work
has been carried out. The financial support by DAAD is gratefully
acknowledged.
\end{acknowledgments}

\begin{appendix}
\appendix*
\section{}
\renewcommand{\theequation}{\thesection.\arabic{equation}}
 \label{sec:appendix}\setcounter{equation}{0}

The components $\varphi _{n}^{1}(u)$ and $\varphi _{n}^{2}(u)$ of the pion two-particle
twist-4 DAs $\varphi_1^{(4)}(u,\mu_F^2)$, $\varphi_2^{(4)}(u,\mu_F^2)$ are given
by the following expressions \cite{A02}:
\[
\varphi _{0}^{1}(u)=\delta ^{2}\left\{ \overline{u}\left[ \ln \overline{u}-%
\rm{Li}_{2}(\overline{u})\right] +u\left[ \ln u-\rm{Li}_{2}(u)\right]
-u\overline{u}+\frac{\pi ^{2}}{6}\right\},
\]
\[
\varphi _{1}^{1}(u)=\delta ^{2}\left\{ \overline{u}\left[ \left( 1+\frac{
\overline{u}}{2}-\frac{\overline{u}^{2}}{3}\right) \ln \overline{u}-\rm{
Li}_{2}(\overline{u})\right] +u\left[ \left( 1+\frac{u}{2}-\frac{u^{2}}{3}
\right) \ln u-\rm{Li}_{2}(u)\right]  
 -\frac{5}{6}u\overline{u}+\frac{1}{2}u^{2}\overline{u}^{2}+\frac{\pi
^{2}}{6}\right\}, 
\]
\[
\varphi _{2}^{1}(u)=\delta ^{2}\left\{ \overline{u}\left[ \left( 1+\overline{
u}-\frac{2}{3}\overline{u}^{2}\right) \ln \overline{u}-\rm{Li}_{2}(%
\overline{u})\right] +u\left[ \left( 1+u-\frac{2}{3}u^{2}\right) \ln u-
\rm{Li}_{2}(u)\right]
 -\frac{2}{3}u\overline{u}+\frac{5}{4}u^{2}\overline{u}^{2}+\frac{\pi
^{2}}{6}\right\}, 
\]
\[
\varphi _{3}^{1}(u)=\delta ^{2}\left\{ \overline{u}\left[ \left( 1+\frac{3}{2%
}\overline{u}-\frac{7}{6}\overline{u}^{2}+\frac{1}{4}\overline{u}^{3}-\frac{1
}{10}\overline{u}^{4}\right) \ln \overline{u}-\rm{Li}_{2}(\overline{u})
\right] \right. 
\]
\[
\left. +u\left[ \left( 1+\frac{3}{2}u-\frac{7}{6}u^{2}+\frac{1}{4}u^{3}-\frac{1}{10}
u^{4}\right) \ln u-\rm{Li}_{2}(u)\right] 
-\frac{31}{60}u\overline{u}+\frac{257}{120}u^{2}\overline{u}^{2}-
\frac{1}{3}u^{3}\overline{u}^{3}+\frac{\pi ^{2}}{6}\right\} ,
\]
\[
\varphi _{4}^{1}(u)=\delta ^{2}\left\{ \overline{u}\left[ \left( 1+2
\overline{u}-\frac{11}{6}\overline{u}^{2}+\frac{3}{4}\overline{u}^{3}-\frac{3
}{10}\overline{u}^{4}\right) \ln \overline{u}-\rm{Li}_{2}(\overline{u})
\right] \right. 
\]
\[
\left. +u\left[ \left( 1+2u-\frac{11}{6}u^{2}+\frac{3}{4}u^{3}-\frac{3}{10}
u^{4}\right) \ln u-\rm{Li}_{2}(u)\right] 
 -\frac{23}{60}u\overline{u}+\frac{47}{15}u^{2}\overline{u}^{2}-\frac{
41}{36}u^{3}\overline{u}^{3}+\frac{\pi ^{2}}{6}\right\},   \label{eq:15}
\]
and
\[
\varphi _{0}^{2}(u)=\delta ^{2}\left[ u^{2}\ln u+\overline{u}^{2}\ln 
\overline{u}+u\overline{u}\right],\;\;
\varphi _{1}^{2}(u)=\delta ^{2}\left[ u^{2}\ln u+\overline{u}^{2}\ln 
\overline{u}+u\overline{u}+\frac{1}{2}u^{2}\overline{u}^{2}\right],
\]
\[
\varphi _{2}^{2}(u)=\delta ^{2}\left[ u^{2}\ln u+\overline{u}^{2}\ln 
\overline{u}+u\overline{u}+\frac{5}{6}u^{2}\overline{u}^{2}\right],\;
\varphi _{3}^{2}(u)=\delta ^{2}\left[ u^{2}\ln u+\overline{u}^{2}\ln 
\overline{u}+u\overline{u}+\frac{13}{12}u^{2}\overline{u}^{2}-\frac{1}{6}
u^{3}\overline{u}^{3}\right],
\]
\[
\varphi _{4}^{2}(u)=\delta ^{2}\left[ u^{2}\ln u+\overline{u}^{2}\ln 
\overline{u}+u\overline{u}+\frac{77}{60}u^{2}\overline{u}^{2}-\frac{13}{30}
u^{3}\overline{u}^{3}\right], 
\]
where ${\rm{Li}}_{a}(x)=\sum_{n=1}^{\infty }x^{n}/n^{a}$.

The twist-4 contribution to the FF (\ref{eq:3.7}) can be formulated in terms of the
twist-4 specrtal density $\rho^{(4)}(Q^2,s)$:
\[
F_{\pi\gamma}^{(4)}(Q^2)=\frac{\sqrt{2}f_{\pi}}{3}\left[\frac{1}{m_{\rho}^2}\int_0^{s_0}ds\rho^{(4)}(Q^2,s)
\exp\left(\frac{m_{\rho}^2-s}{M^2} \right)+\frac{1}{Q^2}H^{(4)}(Q^2) \right ],  
\]
where
\[
H^{(4)}(Q^2)=\int_{s_0}^{\infty}\frac{ds}{s}Q^2\rho^{(4)}(Q^2,s).
\]
The twist-4 spectral density and the function $H^{(4)}(Q^2)$ have the decompositions
\[
\rho^{(4)}(Q^2,s)=2\delta^2(Q^2)\sum_{n=0}^{4}K_n(Q^2)\rho_n^4(Q^2,s)
\]
and
\[
H^{(4)}(Q^2)=\frac{2\delta^2(Q^2)}{Q^2}\sum_{n=0}^4K_n(Q^2)H_n^4(Q^2).
\]
The explicit expressions for the components of $\rho^{(4)}(Q^2,s)$ and $H^{(4)}(Q^2)$
are written down below (hereafter $t\equiv Q^2$):

\[
\rho_0^4(t,s)=\frac{2}{(s+t)^3}\left \{ s-t+2s\ln\left [\frac{s}{s+t}\right]-
2t\ln\left [\frac{t}{s+t}\right ] \right\} ,
\]
\[
\rho_1^4(t,s)=\frac{2}{(s+t)^4}\left \{ s^2-t^2+2s^2\ln\left [\frac{s}{s+t}\right]-
2t^2\ln\left [\frac{t}{s+t}\right ] \right\} ,
\]
\[
\rho_2^4(t,s)=\frac{1}{3(s+t)^5}\left [6s^3+5s^2t-5st^2-6t^3 \right ]+
\frac{2t(s-2t)}{(s+t)^4}\ln \left [\frac{t}{s+t} \right ]+
\frac{2s(2s-t)}{(s+t)^4}\ln \left [\frac{s}{s+t} \right ],
\] 
\[
\rho_3^4(t,s)=\frac{1}{3(s+t)^7}\left [6s^5+19s^4t+10s^3t^2-10s^2t^3-19st^4-6t^5 \right ]
\]
\[
+ \frac{4t(s^3-st^2-t^3)}{(s+t)^6}\ln \left [\frac{t}{s+t} \right ]+
\frac{4s(s^3+s^2t-t^3)}{(s+t)^6}\ln \left [\frac{s}{s+t} \right ],
\]
\[
\rho_4^4(t,s)=\frac{st(s-t)}{30(s+t)^7}\left [21s^2-65st+21t^2 \right ]
\]
\[
+\frac{1}{3(s+t)^6}\left\{ t \left [5s^3-5s^2t-10st^2-6t^3 \right ]+ 
s \left [6s^3+10s^2t+5st^2-5t^3 \right ]
\right\} 
\]
\[
+ \frac{2t(3s^3-s^2t-st^2-2t^3)}{(s+t)^6}\ln \left [\frac{t}{s+t} \right ]+
\frac{2s(2s^3+s^2t+st^2-3t^3)}{(s+t)^6}\ln \left [\frac{s}{s+t} \right ].
\]

Here we introduce new notations
\[
A=\ln \left [\frac{t}{s_0+t}\right ],\; B=\ln \left [\frac{s_0}{s_0+t}\right ],\;\;
S(t)=\frac{2t(2A-1)}{s_0+t}+2B+4\left[\frac{\pi^2}{6}-{\rm {Li}}_2\left (\frac{s_0}{s_0+t} \right )\right],
\]
and get:
\[
H_0^4(t)=\frac{2}{(s_0+t)^2}\left[t^2(A-1)-ts_0-s_0(2t+s_0)B \right]+S(t),
\]
\[
H_1^4(t)=\frac{1}{9(s_0+t)^3}\left[12t^3A-9t^3-ts_0(15t+6s_0)-6s_0^2(3t+s_0)B \right]
+\frac{t^2(1+2A)}{(s_0+t)^2}+S(t),
\]
\[
H_2^4(t)=\frac{t^4}{6(s_0+t)^4}-\frac{t^3}{9(s_0+t)^3}+\frac{2}{(s_0+t)^3}\left[
ts_0(s_0+t)+t^3A+ s_0^3 B\right ]
\]
\[
+\frac{1}{(s_0+t)^2}\left [ t(t-2s_0)+2t^2A+2s_0(t-s_0)B  
\right ]+S(t),
\]
\[
H_3^4(t)=-\frac{t^6}{3(s_0+t)^6}+\frac{1}{5(s_0+t)^5}\left\{ t\left[3t^4+4s_0(t^3+2t^2s_0+2ts_0^2+s_0^3)
\right]+4t^5A +4s_0^5B
\right\}
\]
\[
-\frac{t^4(1+6A)}{6(s_0+t)^4}-\frac{3}{4(s_0+t)^4}
\left [t^4+4t^3s_0+6t^2s_0^2+4ts_0^3+4s_0^4B \right ]
+\frac{t^3(-7+24A)}{9(s_0+t)^3}
\]
\[
+\frac{20}{9(s_0+t)^3}
\left [t^3+3t^2s_0+3ts_0^2+3s_0^3B \right ]
+\frac{t^2(1+2A)}{(s_0+t)^2}-\frac{4s_0}{(s_0+t)^2}
\left [t+(s_0-t)B \right ]+S(t),
\]
\[
H_4^4(t)=-\frac{107t^6}{90(s_0+t)^6}+\frac{t^5(87+100A)}{50(s_0+t)^5}+
\frac{2}{5(s_0+t)^5}\left [t^5+5t^4s_0+10t^3s_0^2+10t^2s_0^3+5ts_0^4+5s_0^5B
\right ]
\]
\[
-\frac{t^4(67+150A)}{60(s_0+t)^4}-
\frac{15}{8(s_0+t)^4}\left [t^4+4t^3s_0+6t^2s_0^2+4ts_0^3+4s_0^4B
\right ]+\frac{t^3(-49+300A)}{90(s_0+t)^3}
\]
\[
+\frac{40}{9(s_0+t)^3}\left [t^3+3t^2s_0+3ts_0^2+3s_0^3B
\right ]
+\frac{1}{2(s_0+t)^2}\left[t(t-14s_0)+4t^2A +2s_0(6t-7s_0)B
\right]+S(t).
\]
\end{appendix}


\begin{thebibliography}{99}
\bibitem{LB80} G.\ P.\ Lepage and S.\ J.\ Brodsky, Phys. Lett. B \textbf{87}%
, 359 (1979); %%CITATION=PHLTA,B87,359%%
Phys. Rev. D \textbf{22}, 2157 (1980). %%CITATION=PHRVA,D22,2157%%
%[1]%

\bibitem{ER80} A.\ V.\ Efremov and A.\ V.\ Radyushkin, Teor. Mat. Fiz. 
\textbf{42}, 147 (1980) [Theor. Math. Phys. \textbf{42}, 97 (1980)]; Phys.
Lett. B \textbf{94}, 245 (1980). %%CITATION=PHLTA,B94,245%%
%[2]%

\bibitem{DM80} A.\ Duncan and A.\ H.\ Mueller, Phys. Rev. D \textbf{21},
1636 (1980). %%CITATION=PHRVA,D21,1636%%
%[3]%

\bibitem{ABR83} F.\ del Aguila and M.\ K.\ Chase, Nucl. Phys. \textbf{B193},
517 (1981);\newline
E.\ Braaten, Phys. Rev. D \textbf{28}, 524 (1983);\newline
E.\ P.\ Kadantseva, S.\ V.\ Mikhailov, and A.\ V.\ Radyushkin, Yad. Fiz. 
\textbf{44}, 507 (1986) [Sov. J. Nucl. Phys. \textbf{44}, 326 (1986)];%
\newline
I.\ V.\ Musatov and A.\ V.\ Radyushkin, Phys. Rev. D \textbf{56}, 2713
(1997);\newline
B.\ Melic, D.\ M\"{u}ller, and K.\ Passek-Kumericki, Phys. Rev. D \textbf{68}%
, 014013 (2003). %%CITATION=PHRVA,D68,014013%%
%[4]%

\bibitem{Kd99} A.\ Khodjamirian, Eur. Phys. J. C \textbf{6}, 477 (1999). 
%%CITATION=EPHJA,6,477%%
%[5]%

\bibitem{SY00} A.\ Schmedding and O.\ Yakovlev, Phys. Rev. D \textbf{62},
116002 (2000). %%CITATION=PHRVA,D62,116002%%
%[6]%

\bibitem{BMS03} A.\ P.\ Bakulev, S.\ V.\ Mikhailov, and N.\ G.\ Stefanis,
Phys. Rev. D \textbf{67}, 074012 (2003). %%CITATION=PHRVA,D67,074012%%
%[7]%

\bibitem{A01} S.\ S.\ Agaev, Phys. Rev. D\ \textbf{69}, 094010 (2004). 
%%CITATION=PHRVA,D69,094010%%
%[8]%

\bibitem{CZ84} V.\ L.\ Chernyak and A.\ R.\ Zhitnitsky, Phys. Rep. \textbf{%
112}, 173 (1984). %%CITATION=PRPLC,112,173%%
%[9]%

\bibitem{Dor} V.\ Yu.\ Petrov, M.\ V.\ Polyakov, R.\ Ruskov, C.\ Weiss, and
K.\ Goeke, Phys. Rev. D \textbf{59}, 114018 (1999);\newline
M.\ Praszalowicz and A.\ Rostworowski, Phys. Rev. D \textbf{64}, 074003
(2001);\newline
A.\ E.\ Dorokhov, Pis'ma Zh. Eksp. Teor. Fiz. \textbf{77}, 68 (2003) [JETP
Lett. \textbf{77}, 63 (2003)]. %%CITATION =HEP-PH 0212156%%
%[10]%

\bibitem{Lattice} T.\ A.\ DeGrand and R.\ D.\ Loft, Phys. Rev. D \textbf{38}%
, 954 (1988);\newline
D.\ Daniel, R.\ Gupta, and D.\ G.\ Richards, Phys. Rev. D \textbf{43}, 3715
(1991);\newline
L.\ Del Debbio, M.\ Di Pierro, and A. Dougall, Nucl. Phys. Proc. Suppl. 
\textbf{119}, 416 (2003);\newline
M.\ G\"ockeler et al., QCDSF/UKQCD Coll., arxiv: hep-lat/0510089. 
%%CITATION =HEP-LAT 0510089%%
%[11]%

\bibitem{BKM03} V.\ M.\ Braun, G.\ P.\ Korchemsky, and D.\ M\"{u}ller, Prog.
Part. Nucl. Phys. \textbf{51}, 311 (2003). %%CITATION = HEP-PH 0306057%%
%[12]%

\bibitem{BF90} V.\ M.\ Braun and I.\ E.\ Filyanov, Z. Phys. C\ \textbf{48},
239 (1990). %%CITATION=ZEPYA,C48,239%%
%[13]%

\bibitem{BB98} P.\ Ball, V.\ M.\ Braun, Y.\ Koike, and K.\ Tanaka, Nucl.
Phys. \textbf{B529}, 323 (1998);\newline
P. Ball, V.\ M.\ Braun, Nucl. Phys. \textbf{B543}, 201 (1999);\newline
P. Ball, J. High Energy Phys. \textbf{01}, 010 (1999). 
%%CITATION = HEP-PH 9812375%%
%[14]%

\bibitem{An00} J. R. Andersen, Phys. Lett. B \textbf{475}, 141 (2000). 
%%CITATION=PHLTA,B475,141%%
%[15]%

\bibitem{BGG04} V.\ M.\ Braun, E.\ Gardi, and S.\ Gottwald, Nucl. Phys. 
\textbf{B685}, 171 (2004). %%CITATION=NUPHA,B685,171%%
%[16]%

\bibitem{A02} S.\ S.\ Agaev, Phys. Rev. D \textbf{72}, 074020 (2005). 
%%CITATION=PHRVA,D72,074020%%
%[17]%

\bibitem{BH94} V. Braun and I. Halperin, Phys. Lett. B\ \textbf{328}, 457
(1994);\newline
V. M. Braun, A. Khodjamirian, and M. Maul, Phys. Rev. D\ \textbf{61}, 073004
(2000);\newline
J. Bijnens and A. Khodjamirian, Eur. Phys. J C \textbf{26}, 67 (2002). 
%%CITATION=EPHJA,26,67%%
%[18]%

\bibitem{CELLO} H.-J.\ Behrend \textit{et al}. (CELLO Collaboration), Z.
Phys. C \textbf{49}, 401 (1991). %%CITATION=ZEPYA,C49,401%%
%[19]%

\bibitem{CLEO} J.\ Gronberg \textit{et al}. (CLEO Collaboration), Phys. Rev.
D \textbf{57}, 33 (1998). %%CITATION=PHRVA,D57,33%%
%[20]%

\bibitem{Br89} I. I. Balitsky and V. M. Braun, Nucl. Phys. \textbf{B311},
541 (1989). %%CITATION=NUPHA,B311,541%%
%[21]%

\bibitem{NS94} V. L. Chernyak, A. R. Zhitnitsky, and I. R. Zhitnitsky, Yad.
Fiz. \textbf{38}, 1074 (1983) [ Sov. J. Nucl. Phys. \textbf{38}, 645 (1983)];%
\newline
V. A. Novikov, M. A. Shifman, A. I. Vainstein, M. B. Voloshin, and V. I.
Zakharov, Nucl. Phys. \textbf{B237}, 525 (1984). 
%%CITATION=NUPHA,B237,525%%
%[22]%

\bibitem{DRS} F.\ M.\ Dittes and A.\ V.\ Radyushkin, Phys. Lett. B \textbf{%
134}, 359 (1984);\newline
M.\ H.\ Sarmadi, Phys. Lett. B \textbf{143}, 471 (1984);\newline
S.\ V.\ Mikhailov and A.\ V.\ Radyushkin, Nucl. Phys. \textbf{B254}, 89
(1985). %%CITATION=NUPHA,B254,89%%
%[23]%

\bibitem{BBK89} I. I. Balitsky, V. M. Braun, and A. V. Kolesnichenko, Nucl.
Phys. \textbf{B312}, 509 (1989);\newline
V. M. Braun and I. E. Filyanov, Z. Phys. C\ \textbf{44}, 157 (1989);\newline
V. L. Chernyak and I. R. Zhitnitsky, Nucl. Phys. \textbf{B345}, 137 (1990). 
%%CITATION=NUPHA,B345,137%%
%[24]%

\bibitem{BMS05} A.\ P.\ Bakulev, S.\ V.\ Mikhailov, and N.\ G.\ Stefanis,
arxiv: hep-ph/0512119. %%CITATION=HEP-PH 0512119%%
%[25]%

\bibitem{SVZ} M. A. Shifman, A. I. Vainstein, and V. I. Zakharov, Nucl.
Phys. \textbf{B147}, 385; \textbf{B147} 448 (1979). 
%%CITATION=NUPHA,B147,385%%
%%CITATION=NUPHA,B147,448%%
%[26]%
\end{thebibliography}
\end{document}